\documentclass[a4paper, 12pt, titlepage]{article}
\usepackage[english]{babel}
\usepackage[latin1]{inputenc}   
\usepackage{times}
\usepackage{a4wide}             
\usepackage{fancyhdr}
\usepackage{amssymb}
\usepackage{latexsym}
\usepackage{epic}
\usepackage{eepic}
\usepackage{graphics}
\usepackage[T1]{fontenc}
\usepackage{graphicx}   
\usepackage{epsfig}     
\usepackage{here}
\usepackage[hang,small,bf]{caption}

\begin{document}

\begin{center}

\Large{\bf{Measurement as Soft Final-State Interaction\\ with a
  Stochastic System}}

\vspace{1 cm}

\large{June 2002\\
K.-E. Eriksson, Division of Engineering Sciences, Physics and Mathematics\\
  Karlstad University,\\
 SE 65188 Karlstad, Sweden\\ Karl-Erik.Eriksson@kau.se}
\end{center}

\normalsize{ }

\vspace{3 cm}

\noindent
\large{\textbf{Abstract}}

\noindent
\normalsize{
A small quantum scattering system (the microsystem) is studied in interaction
with a large quantum system (the macrosystem)  described by unknown stochastic
variables. The interaction between the two systems is diagonal for the
microsystem in a certain basis, and it leads to an imprint on the macrosystem.
Moreover, the interaction is assumed to involve only small transfers of energy
and momentum between the two systems (as compared to typical energies/momenta
within the microsystem). This makes it suitable to carry out the analysis in
scattering theory, where the transition amplitude for the whole system
factorizes.
The interaction taking place within the macrosystem
is assumed to depend on the stochastic variables in such a way that, on the
average, no particular channel is favoured.
The result is then, in the thermodynamic limit of the macrosystem,
that the
whole system bifurcates and the microsystem
 ends up in a state described by one of the basis vectors (in the
mentioned basis). The macrosystem ends up in an entangled state tied to this
basis vector.
For the ensemble of macrosystems, the interaction with the microsystem
leads, on
the average, to the usual decoherence and  diagonal density matrix for the
microsystem.
The macrosystem can be interpreted as representing a measurement device for
performing a measurement on the microsystem.
The whole discussion is carried out within quantum mechanics itself
without any
modification or generalization.}

\vspace{5 cm}
\thispagestyle{empty}

\pagebreak

\section{Introduction: the problem}

The aim of this paper is to analyse a microscopic quantum event in a
microscopic quantum system $\mu$  \textit{together} with a related
interaction with a
macroscopic system $M$, not known in any detail and therefore described
by stochastic variables. The intention is to model a measurement
process, where $M$ is a measurement device for performing a measurement
on $\mu$.

Let us assume that an observable $A$ with non-degenerate eigenstates
$|j\rangle_\mu$ is
to be measured. The interaction between $\mu$  and $M$  must then be such that
the state $|j\rangle_\mu$ of $\mu$  makes an imprint on $M$. We
assume this to take place
without the state of  $\mu$ being changed.

In the discussion of measurement it has often been assumed that the process
within $\mu$  can be analysed independently of the interaction between
$\mu$  and $M$ (assumption X), although the final state of $\mu$ is
registered only through measurement, i.e., only through the
interaction between $\mu$ and $M$.

If the interaction between $\mu$ and $M$ is analysed under the assumption X
of a given final state from the process within $\mu$, and if the
interaction between $\mu$ and $M$ is of the kind just described, then in
general, a reduction of the state of $\mu$  into an eigenstate of $A$ cannot take
place without violating the superposition principle \cite{1}. This situation
has led to a common notion that there may be two kinds of interaction,
(i) a microscopic interaction within the system $\mu$ itself,
obeying a linear Schr\"{o}dinger equation, and (ii) another type of
interaction between $\mu$ and $M$. A possibility to combine both of these
effects is to have a non-linear equation of motion 
\cite{2} to \cite{5}, where the
non-linearities become non-negligible only in interactions involving
mesoscopic or macroscopic objects.

In this paper we show that a non-linearity arises within quantum mechanics
itself through extension of the system considered, to include also $M$.

All since the famous Einstein-Podolski-Rosen paper \cite{6}, quantum
entanglement
phenomena  have shown consequences that have been regarded as
counter-intuitive. In this analysis, we use S-matrix theory to investigate the
rôle of quantum entanglement between $\mu$  and $M$  through a final-state interaction,
that gives a factor in the overall transition matrix.

The point of view taken here, not accepting the assumption X, is that
it is necessary to analyse the interaction within $\mu$ and the
interaction between $\mu$ and $M$ as a whole. As pointed out already,
the interaction between $\mu$ and $M$ is assumed to
give $M$ an imprint from $\mu$ without changing the eigenstate of the
observable $A$ for $\mu$.
It then also leads to an entanglement of $\mu$ with the (metastable)
system $M$. The
stochastic variables characterizing $M$ may have an enhancing or inhibiting
influence on the
transitions within $M$ to a final state. Therefore, the different
initial states of
the metastable system $M$, described by stochastic variables, compete
on an unequal
basis to reach the final state, and the ensemble of final states can
have a very
different composition from that of the initial states.

The system $M$ should not only be metastable; it should also be
 unbiased. We take
this to mean that the corresponding enhancement factors and inhibition factors
of $M$ occur with the same frequency in the initial state.

Since we are studying the process of internal interaction and measurement as a
whole, it is natural to conduct the analysis within the framework of
 scattering
theory. Moreover, in the limit of low energy and momentum transfer,
the $\mu$-$M$
interaction factorizes in the scattering amplitude (see Appendix 1) and hence
also in the transition probability per unit time. The factor from
$\mu$-$M$
interaction
depends on $\mu$ only through its outgoing state.

The stochastic variables of $M$ are introduced through a stepwise mapping
procedure, thus going from the situation of the microsystem $\mu$ by itself
to a situation where $\mu$ interacts with the system $M$ in the thermodynamic
limit, i.e., in the limit of an infinite number of stochastic
variables.

This mapping is non-linear, and the non-linearity can be understood in the
formalism of perturbation theory, where the internal $\mu$ interaction
has to appear
in Feynman diagrams mixed with the $\mu$-$M$ interaction (Appendix 2).

This mapping turns out to be a bifurcation process, describable in the
probability simplex of $\mu$ as a random walk (brownian motion), ending
up in one of
the corners. The ensemble of such walks is then a diffusion process with the
corners of the simplex as attractors.

A random-walk process is not new in this context. It results from a non-linear
dynamics like the one suggested in \cite{5}. It must be emphasized
that the model in this paper
is an S-matrix model not describing the detailed dynamics. As
mentioned already,
the mapping here interpolates between a situation without the system $M$ and a
situation with the system $M$ where $M$ has many degrees
of freedom (the thermodynamic limit).

In the next section, the scattering process is described for the
situation of $\mu$
without $M$. The modifications due to a final-state interaction with
$M$ is described
in Section 3. The stochastic variables are introduced in Section 4, and their
influence on the whole process is taken into account. Section 5 describes the
mapping procedure, the resulting random-walk or diffusion process and the
thermodynamic limit of $M$. In Section 6, a simplified model is shown
in detail. In
Section 7, we show, how correlations between stochastic variables of
$M$ can build
up through entanglement with $\mu$. In Section 8, we indicate how the measurement
process can be interpreted as an evolution process in Darwinian terms. Finally, we state briefly
a few conclusions in the last section.

\section{The underlying scattering process}

We assume that the microscopic quantum system $\mu$ has an internal
dynamics taking it from an initial state  $|0\rangle_{\mu}$, which is
an eigenstate of
the free Hamiltonian $H_{0}$,
\begin{equation}
H_{0}|0\rangle_{\mu}=E_{0}|0\rangle_{\mu},~~~~_{\mu}\langle 0|0\rangle_{\mu}=1,
\end{equation}
to a certain final state  $|\Psi\rangle_{\mu}$, which is also an
eigenstate of $H_{0}$, but
different from the initial state (for instance localized in a
different spatial region),
\begin{equation}
H_{0}|\Psi\rangle_{\mu}=E|\Psi\rangle_{\mu},~~~~_{\mu}\langle
\Psi |\Psi\rangle_{\mu}=1,~~~~_{\mu}\langle 0|\Psi\rangle_{\mu}=0.
\end{equation}
Consider an orthonormal basis for the final states,
\begin{equation}
|j\rangle_{\mu},~~~j=1,2,...,n;~~~~_{\mu}\langle j|0\rangle_{\mu}=0;
~~~~_{\mu}\langle j|k\rangle_{\mu}=\delta_{jk},
\end{equation}
formed by states that are degenerate eigenstates of $H_{0}$  and
non-degenerate eigenstates of an observable $A=A^{\dagger}$ acting on $\mu$,
\begin{equation}
\begin{array}{l}
H_{0}|j\rangle_{\mu}=E|j\rangle_{\mu};\\
A|j\rangle_{\mu}=a_{j}|j\rangle_{\mu},~~~a_{j}^{\ast}=a_{j},~~~a_{j}\neq
a_{k}~\mbox{for}~j\neq k;\\
A|0\rangle_{\mu}=a_{0}|0\rangle_{\mu},~~~a_{0}^{\ast}=a_{0},~~~a_{0}\neq
a_{k}.
\end{array}
\end{equation}
Let the final state in this basis be
\begin{equation}
|\Psi\rangle_{\mu}=\sum_{j=1}^{n}\Psi_{j}|j\rangle_{\mu};~~~~
\sum_{j=1}^{n}|\Psi_{j}|^{2}=1.
\end{equation}
Because of (4), the observable $A$ commutes with $H_{0}$, and it is
suitable for
describing an outgoing state. Clearly the internal
interaction Hamiltonian for  $\mu$, $H_{I}$, does not commute with $A$,
\begin{equation}
[H_{I},A]\neq 0.
\end{equation}
Then the S-matrix elements are proportional to the components of
$|\Psi\rangle_{\mu}$ in (5),
\begin{equation}
_{\mu}\langle j|S|0\rangle_{\mu}=~_{\mu}\langle
j|M|0\rangle_{\mu}~\delta(E-E_{0}),
\end{equation}
with
\begin{equation}
_{\mu}\langle j|M|0\rangle_{\mu}=\sqrt{\Gamma}\Psi_{j}.
\end{equation}
This is the transition probability amplitude for a transition from the
initial state $|0\rangle_{\mu}$  to the final state $|j\rangle_{\mu}$
of $\mu$.

The transition probability per unit time for this transition is then
\begin{equation}
(2\pi)^{-1}|_{\mu}\langle j|M|0\rangle_{\mu}|^{2}~\delta(E-E_{0})=
(2\pi)^{-1}\Gamma|\Psi_{j}|^{2}~\delta(E-E_{0}),
\end{equation}
and the total transition probability per unit time for a transition to
any of the final states is
\begin{equation}
(2\pi)^{-1}\Gamma\delta(E-E_{0}).
\end{equation}
The density operator of the initial state for given energy is
\begin{equation}
\rho_{0}=|0\rangle_{\mu}~_{\mu}\langle 0|;~~~\mbox{Tr}\rho_{0}=1.
\end{equation}
The corresponding density operator of the final state is
\begin{equation}
\rho_{f}=\Gamma^{-1}M|0\rangle_{\mu}~_{\mu}\langle
0|M^{\dagger}~=~|\Psi\rangle_{\mu}~_{\mu}\langle
\Psi|~=~\sum_{j,k=1}^{n}\Psi_{j}\Psi_{k}^{\ast}|j\rangle_{\mu}~_{\mu}\langle
k|;~~~\mbox{Tr}\rho_{f}=1.
\end{equation}

\section{Changes due to soft final-state interaction}

We now introduce the interaction between the particles in the outgoing
states (the eigenstates of $A$) of $\mu$ and the large system $M$. We
assume this
interaction to involve only very small momentum and energy
transfers. As pointed out already, the interaction between $\mu$ and $M$ and
within $M$ is assumed to be such that an eigenstate $|j\rangle_{\mu}$ of $A$
for $\mu$ does not change but
makes an imprint on $M$. We assume the initial states of $M$ to be
$|0;\underline{\underline{\varepsilon}}\rangle_{M}$, where '$0$'
denotes 'no imprint' and where $\underline{\underline{\varepsilon}}$ is a set of
stochastic variables to
describe the (unknown) structure of $M$. We assume the variables
$\underline{\underline{\varepsilon}}$ to be so defined that they are constants of motion.
The corresponding final states
can then be denoted by $|j;\underline{\underline{\varepsilon}}\rangle_{M}$, where
'$j$' denotes an imprint on $M$.

Here the assumption is made that the measurement interaction copies
the state of $\mu$ without changing it. For the class of measurements where
the state of $\mu$ is changed, a generalized formalism is needed. Such a
generalization seems to be rather straightforward however.

Thus the initial state of the combined system $\mu+M$ is
\begin{equation}
|0\rangle_{\mu}\otimes|0;\underline{\underline{\varepsilon}}\rangle_{M},
\end{equation}
and a corresponding set of final states are
\begin{equation}
|j\rangle_{\mu}\otimes|j;\underline{\underline{\varepsilon}}\rangle_{M}.
\end{equation}
The soft (low momentum-transfer) character of the $\mu$-$M$ interaction
implies a factorization of this interaction. This is well-known and has been known for a long time
\cite{7},
but we show the factorization for soft electromagnetic interaction in
Appendix 1. Thus, the element of the new $M$-matrix denoted by
$\overline{M}$, is then a modified version of (8),
\begin{equation}
\left(~_{\mu}\langle j|\otimes~_{M}\langle
j;\underline{\underline{\varepsilon}}|\right)
\overline{M}\left(|0\rangle_{\mu}\otimes|0;\underline{\underline{\varepsilon}}\rangle_{M}
\right)~=~b_{j}(\underline{\underline{\varepsilon}})\sqrt{\Gamma}\Psi_{j}.
\end{equation}
The factor $b_{j}(\underline{\underline{\varepsilon}})$  gives the modification
due to final-state interaction, to
be discussed in more detail below.

This means that for a given set of stochastic variables
$\underline{\underline{\varepsilon}}$  the final
state of the combined system, corresponding to (5) is proportional to
\begin{equation}
\overline{M}\left(|0\rangle_{\mu}\otimes|0;\underline{\underline{\varepsilon}}\rangle_{M}
\right)=\sqrt{\Gamma}\sum_{j=1}^{n}b_{j}(\underline{\underline{\varepsilon}})\Psi_{j}|j
\rangle_{\mu}\otimes|j;\underline{\underline{\varepsilon}}\rangle_{M},
\end{equation}
which, in general, is not normalized.

In any scattering process, there is emission of soft radiation of
photons and gravitons, which factorizes like in the description given
here. Since this radiation is not detected, a summation over final
states is necessary. In (9) this means an integration over how the
energy $E_{0}$ is shared by the soft radiation and the other particles
of $\mu$. We can leave out the $\delta$-function, and use (11) and (12) for
the
density operator, which also makes normalization simpler. The energy
$\delta$-function for the integrated process involving $\mu$ and $M$
is $\delta(E+\Delta E-E_{0})$, where $E$ is
the final energy of $\mu$, and where $\Delta E$ is the (very small)
energy transfer
from $\mu$ to $M$. Thus $E$ is approximately equal to $E_{0}$.

The equations for the density operators corresponding to (11) and
(12), are now
\begin{equation}
\rho_{0}(\underline{\underline{\varepsilon}})=(|0\rangle_{\mu}\otimes|0;
\underline{\underline{\varepsilon}}\rangle_{M})(_{\mu} \langle
0|\otimes~ _{M}\langle 0;\underline{\underline{\varepsilon}}|)
\end{equation}
for the initial state, and
$$
\rho_{f}(\underline{\underline{\varepsilon}})=\overline{\Gamma}^{-1}\overline{M}
\rho_{0}(\underline{\underline{\varepsilon}})\overline{M}^{\dagger}=$$
\begin{equation}
=\left(\sum_{l=1}^{n}|b_{l}(\underline{\underline{\varepsilon}})|^{2}|\Psi_{l}
|^{2}\right)^{-1}\sum_{j,k=1}^{n}b_{j}(\underline{\underline{\varepsilon}})b_{k}
(\underline{\underline{\varepsilon}})^{\ast}\Psi_{j}\Psi_{k}^{\ast}(|j
\rangle_{\mu}\otimes|j;\underline{\underline{\varepsilon}}\rangle_{M})(_{\mu}\langle
k|\otimes~ _{M}\langle k;\underline{\underline{\varepsilon}}|)
\end{equation}
for the final state, where
\begin{equation}
\overline{\Gamma}(\underline{\underline{\varepsilon}})=\Gamma\sum_{l=1}^{n}|b_{l}
(\underline{\underline{\varepsilon}})|^{2}|\Psi_{l}|^{2}
\end{equation}
is the new probability per unit time for a transition.

We notice that the density operator (18) is non-linear in
$|\Psi_{j}|^{2}$. This is
quite natural, since in the total scattering operator, the internal
interaction in  $\mu$, described by the S-matrix elements given by (7) and
(8), and the  $\mu$-$M$ interaction appear mixed to all orders, and a
perturbation expansion would show these non-linearities as
diagrammatically described in Appendix 2.  However, due
to the factorization (16), the internal  $\mu$-interaction only appears
through its  $M$-matrix elements (8).




\section{The stochastic variables and their influence}

For simplicity we assume that the stochastic variables describing $M$
are sign factors,
\begin{equation}
\underline{\underline{\varepsilon}}=(\varepsilon_{jx});~~\varepsilon_{jx}=±1;~~
j=1,2,...,n;~~x=1,2,...,X~\mbox{where}~X>>1.
\end{equation}
and that they contribute to $b_{j}(\underline{\underline{\varepsilon}})$ with
random enhancement/inhibition factors
\begin{equation}
b_{j}(\underline{\underline{\varepsilon}})=G\prod_{x=1}^{X}\prod_{l=1}^{n}\left(
1+\eta_{lx}\left(\delta_{lj}-\frac{1}{2}\right)\varepsilon_{lx}\right);
~~~\eta_{lx}^{\ast}=\eta_{lx},~~~0<\eta_{lx}<<1,
\end{equation}
which can be ascribed to  $\underline{\underline{\varepsilon}}$-dependence in the
interactions within $M$. Then, to second order in the $\eta$'s,
$$
|b_{j}(\underline{\underline{\varepsilon}})|^{2}=\chi
B_{j}(\underline{\underline{\varepsilon}});$$
\begin{equation}
B_{j}(\underline{\underline{\varepsilon}})=
\prod_{x=1}^{X}\prod_{l=1}^{n}\left(1+\eta_{lx}\left(2\delta_{lj}-1\right)
\varepsilon_{lx}\right);~~~\chi=|G|^{2}\prod_{x=1}^{X}\prod_{l=1}^{n}
\left(1+\frac{1}{4}\eta_{lx}^{2}\right).
\end{equation}
Moreover, we assume the ensemble of incoming states of $M$ to be
unbiased. Thus, all values of  $\underline{\underline{\varepsilon}}$ are equally probable
initially, each
with a probability 2$^{-nX}$. Due to the  $\underline{\underline{\varepsilon}}$-dependent
enhancement/inhibition
factors, in the ensemble of final states, however, the transition
probabilities
differ. The probability for a transition taking place from an initial
state, where $M$ is described by  $\underline{\underline{\varepsilon}}$ is
\begin{equation}
P(\underline{\underline{\varepsilon}})=\frac{\overline{\Gamma}(\underline{\underline{\varepsilon}})}
{\sum_{\underline{\underline{\varepsilon}}'}\overline{\Gamma}(\underline{\underline{\varepsilon}}')}
=
2^{-nX}\sum_{j=1}^{n}|\Psi_{j}|^{2}B_{j}(\underline{\underline{\varepsilon}});~~~
\sum_{\underline{\underline{\varepsilon}}}P(\underline{\underline{\varepsilon}})=1.
\end{equation}
The final-state density operator (18) can now be written
\begin{equation}
\rho_{f}(\underline{\underline{\varepsilon}})=\left(\sum_{l=1}^{n}|\Psi_{l}|^{2}B_{l}
(\underline{\underline{\varepsilon}})\right)^{-1}\sum_{j,k=1}^{n}\sqrt{B_{j}
(\underline{\underline{\varepsilon}})B_{k}(\underline{\underline{\varepsilon}})}\Psi_{j}
\Psi_{k}^{\ast}(|j\rangle_{\mu}\otimes
|j;\underline{\underline{\varepsilon}}\rangle_{M})(_{\mu}\langle k|\otimes~
_{M}\langle k;\underline{\underline{\varepsilon}}|).
\end{equation}
With the probability distribution (23), the average over $M$ and
 $\underline{\varepsilon}$ of (24) is
$$
\langle\rho_{\mu
f}(\underline{\underline{\varepsilon}})\rangle_{\underline{\underline{\varepsilon}}}=
\sum_{\underline{\underline{\varepsilon}}}P(\underline{\underline{\varepsilon}})\mbox{Tr}_{M}
[\rho_{f}(\underline{\underline{\varepsilon}})]=$$
\begin{equation}
=2^{-nX}\sum_{\underline{\underline{\varepsilon}}}\sum_{j,k=1}^{n}\sqrt{B_{j}
(\underline{\underline{\varepsilon}})B_{k}(\underline{\underline{\varepsilon}})}\Psi_{j}
\Psi_{k}^{\ast}|j\rangle_{\mu}~_{\mu}\langle k|=\sum_{j=1}^{n}|\Psi_{j}|^{2}
|j\rangle_{\mu} ~_{\mu}\langle j|+\theta_{\mu}.
\end{equation}
Here $\theta_{\mu}\longrightarrow 0$ in the
limit of large $X$, since, to second order in the $\eta$'s,
$$
2^{-nX}\sum_{\underline{\underline{\varepsilon}}}\sqrt{B_{j}(\underline{\underline{\varepsilon}})
B_{k}(\underline{\underline{\varepsilon}})}=$$
$$
=2^{-nX}\sum_{\underline{\underline{\varepsilon}}}\prod_{x=1}^{X}\prod_{l=1}^{n}\left(
1+\eta_{lx}\left(\delta_{lj}+\delta_{lk}-1\right)\varepsilon_{lx}-
\frac{1}{2}\eta_{lx}^{2}(\delta_{lj}+\delta_{lk}-2\delta_{lj}\delta_{lk})
\right)=$$
\begin{equation}
=\delta_{jk}+(1-\delta_{jk})\prod_{x=1}^{X}\left(1-\frac{1}{2}
\eta_{jx}^{2}-\frac{1}{2}\eta_{kx}^{2}\right),
\end{equation}
which goes to $\delta_{jk}$ in the limit of large $X$. Thus, in the
thermodynamical limit of $M$,
\begin{equation}
\langle\rho_{\mu
f}(\underline{\underline{\varepsilon}})\rangle_{\underline{\underline{\varepsilon}}}=\sum_{j=1}^{n}
|\Psi_{j}|^{2}|j\rangle_{\mu}~_{\mu}\langle j|,
\end{equation}
i.e., in the ensemble of final states, the mean of the density matrix
for $\mu$  becomes diagonal with the elements $|\Psi_{j}|^{2}$. This
is the standard
decoherence result for averaging over the ensemble of mesurements.

In the next section, we shall study the development of the density
matrix itself,
\begin{equation}
\rho_{\mu f}(\underline{\underline{\varepsilon}})=\mbox{Tr}_{M}[\rho_{f}
(\underline{\underline{\varepsilon}})]
\end{equation}
for varying  $\underline{\underline{\varepsilon}}$, i.e., we shall study the whole
ensemble
of measurement processes, not just ensemble averages.

\section{A mapping procedure in the stochastic variables}

To go from the situation with only $\mu$ present to the situation with a
system $M$ in its thermodynamical limit, we shall use a mapping procedure
where the $X$th step goes from $n(X-1)$ to $nX$ stochastic variables. We let
$X$ increase from 0 to a large value, $X>>1$.

We then get a recursive equation for the elements of the density matrix (24),
$$
(\rho_{fX})_{jk}=\left(\sum_{l=1}^{n}|\Psi_{l}|^{2}B_{l}(\underline
{\varepsilon}_{X})\right)^{-1}\sqrt{B_{j}(\underline{\varepsilon}_{X})
B_{k}(\underline{\varepsilon}_{X})}\Psi_{j}\Psi_{k}^{\ast}=
$$
\begin{equation}
=(\rho_{f(X-1)})_{jk}\frac{\left(1+\eta_{jX}\varepsilon_{jX}+
\eta_{kX}\varepsilon_{kX}-\sum_{l=1}^{n}\eta_{lX}\varepsilon_{lX}\right)
\left(1-\frac{1}{2}(1-\delta_{jk})(\eta_{jX}^{2}+\eta_{kX}^{2})\right)}
{1+2\sum_{l'=1}^{n}(\rho_{f(X-1)})_{l'l'}\eta_{l'X}\varepsilon_{l'X}-
\sum_{l'=1}^{n}\eta_{l'X}\varepsilon_{l'X}}.
\end{equation}
Thus, the change in the density matrix in this step of the mapping is
\[ \Delta (\rho_{fX})_{jk} = (\rho_{fX})_{jk}-(\rho_{f(X-1)})_{jk}\]
\begin{equation}
 = (\rho_{f(X-1)})_{jk}\frac{\eta_{jX}\varepsilon_{jX}+\eta_{kX}\varepsilon_{kX}-2
\sum_{l=1}^{n}(\rho_{f(X-1)})_{ll}\eta_{lX}\varepsilon_{lX}-\frac{1}{2}
(1-\delta_{jk})(\eta_{jX}^{2}+\eta_{kX}^{2})}
{1+2\sum_{l'=1}^{n}(\rho_{f(X-1)})_{l'l'}\eta_{l'X}\varepsilon_{l'X}-
\sum_{l'=1}^{n}\eta_{l'X}\varepsilon_{l'X}}.
\end{equation}
The probability for a certain
\begin{equation}
\underline{\varepsilon}_{X}=(\varepsilon_{jX});~~~j=1,2,...,n
\end{equation}
in the $X$th step, given all previous steps, is
\begin{equation}
2^{-n}\left(1+2\sum_{l'=1}^{n}(\rho_{f(X-1)})_{l'l'}\eta_{l'X}
\varepsilon_{l'X}-\sum_{l'=1}^{n}\eta_{l'X}\varepsilon_{l'X}\right).
\end{equation}
For the average values and the variation of (28) we get, suppressing
subscripts $f$ and $X$ \cite{4},
$$
\langle\Delta\rho_{jk}\rangle=-\frac{1}{2}(1-\delta_{jk})(\eta_{j}^{2}+
\eta_{k}^{2})\rho_{jk};
$$
$$
\langle\Delta\rho_{jk}\Delta\rho_{lm}\rangle=\rho_{jk}\rho_{lm}\mbox{
\Huge{$\left(\right.$}}\delta_{jl}\eta_{j}\eta_{l}+\delta_{jm}\eta_{j}
\eta_{m}+\delta_{kl}\eta_{k}\eta_{l}+\delta_{km}\eta_{k}\eta_{m}-
$$
\begin{equation}
-2(\rho_{jj}\eta_{j}^{2}+\rho_{kk}\eta_{k}^{2}+\rho_{ll}\eta_{l}^{2}+
\rho_{mm}\eta_{m}^{2})+4\sum_{s=1}^{n}\rho_{ss}^{2}\eta_{s}^{2}\mbox{
\Huge{$\left.\right)$}}.
\end{equation}
The first equation here describes a drift of the non-diagonal elements
to zero. For the diagonal elements in the probability simplex,
$$
\underline{p}=(p_{1},p_{2},...,p_{n});
$$
\begin{equation}
p_{j}=\rho_{jj}>0,~~~\sum_{k=1}^{n}p_{k}=1,
\end{equation}
we get the following random walk [4,8], 
$$
\langle\Delta p_{j}\rangle=0;
$$
\begin{equation}
\langle\Delta p_{j}\Delta p_{k}\rangle=4p_{j}p_{k}\left(\delta_{jk}
\eta_{j}^{2}-p_{j}\eta_{j}^{2}-p_{k}\eta_{k}^{2}+\sum_{l=1}^{n}p_{l}^{2}
\eta_{l}^{2}\right).
\end{equation}
Going to the case of a continuous (suitably normalized) step variable $X$
for the ensemble of random walks, this leads to the following
diffusion equation [4,8] 
in the simplex (34),
\begin{equation}
\frac{\partial}{\partial
  X}F(\underline{p},X)=\frac{1}{2}\frac{\partial^{2}}{\partial
  p_{j}\partial p_{k}}\left( p_{j}p_{k}\left(\delta_{jk}
\eta_{j}^{2}-p_{j}\eta_{j}^{2}-p_{k}\eta_{k}^{2}+\sum_{l=1}^{n}
p_{l}^{2}\eta_{l}^{2}\right)F(\underline{p},X)\right),
\end{equation}
where we have introduced the density function for the ensemble in the simplex,
$$
F(\underline{p},X)\geq 0~~~~\int d\underline{p} F(\underline{p},X)=1;
$$
\begin{equation}
F(\underline{p},0)=\delta(\underline{p};\underline{p}_{\Psi}); ~~~~
\underline{p}_{\Psi}=\left(|\Psi_{1}|^{2},~|\Psi_{2}|^{2},~...~|\Psi_{n}
|^{2}\right).
\end{equation}
Here $d\underline{p}$ is the ($n-1$)-dimensional normalized volume
element of the simplex, and $\delta(\underline{p};\underline{p}_{0})$ is
the corresponding $\delta$-function.

The entropy function in the probability simplex,
\begin{equation}
S(\underline{p})=-\sum_{j=1}^{n} p_{j}\ln p_{j}
\end{equation}
has an $X$-dependent mean value
\begin{equation}
\overline{S}(X)=\int d\underline{p} F(\underline{p},X) S(\underline{p}).
\end{equation}
The diffusion equation in $X$ (36) implies
$$
\frac{\partial}{\partial X}\overline{S}(X)=$$
$$=\frac{1}{2}\int d\underline{p}
F(\underline{p},X)\sum_{j,k=1}^{n}\frac{\partial^{2}S(\underline{p})}{\partial
  p_{j}\partial p_{k}}p_{j}p_{k}\left(\delta_{jk}\eta_{j}^{2}-p_{j}
\eta_{j}^{2}-p_{k}\eta_{k}^{2}+\sum_{l=1}^{n}p_{l}^{2}\eta_{l}^{2}\right)=$$
\begin{equation}
=-\int d\underline{p}
F(\underline{p},X)\sum_{j=1}^{n}\eta_{j}^{2}p_{j}(1-p_{j})\leq 0,
\end{equation}
which means that the diffusion continues with decreasing entropy,
asymptotically approaching a distribution over states with
\begin{equation}
p_{j}(1-p_{j})= 0~~~\mbox{all} j,
\end{equation}
i.e., a distribution with support only in the corners of the
probability simplex
 [4,8]. 
Thus a random walk of repeated mappings (35) ends up in one of the corners.

As can be shown from (36), similar to (40), the mean of $\underline{p}$,
\begin{equation}
\overline{p_{j}}(X)=\int d\underline{p}F(\underline{p},X)p_{j}
\end{equation}
stays constant,
\begin{equation}
\frac{\partial}{\partial X}\overline{p_{j}}(X)=0
\end{equation}
in agreement with (35). Thus, the probability to reach the $m$th
corner, $p_{m}=1$, is given by the original probability before the mappings,
for $X=0$, i.e., $|\Psi_{m}|^{2}$. Let $\delta_{m}(\underline{p})$ be
the $\delta$-function for the $m$th corner. Then
\begin{equation}
F(\underline{p},\infty)=\sum_{m=1}^{n}|\Psi_{m}|^{2}\delta_{m}(\underline{p}).
\end{equation}
For the ensemble of mapping procedures, the chain of mappings is thus a
diffusion process starting from the unimodal distribution in (37) and
bifurcating into the $n$-modal distribution (44). For the single case, i.e., a
single mesurement, the procedure always ends up in one of the corners of the
probability simplex for $\mu$, that means in an eigenstate of $A$.

\section{A simple model}

In this section we give a simple model with a two-state microsystem as an
example where the mathematics can be carried out in detail.

Let us go back to (23) and (24) with  $n=2$ and
$\varepsilon_{1x}=-\varepsilon_{2x}(x=1,2,...,X)$. Furthermore, we let
all $\eta_{lx}$
have the same value $\frac{1}{2}\eta$. Let $X_{j}$ be the number of
$\varepsilon_{jx}$ (varying $x$) that are equal
to  +1 ($X_{1}+X_{2}=X$).
There are $\frac{X!}{X_{1}!X_{2}!}$ $\underline{\varepsilon}_{X}$'s of this kind
and with this constellation, the probability for a
transition is, according to (23) and (22),
\begin{equation}
P(X_{1},X_{2})=\frac{X!}{X_{1}!X_{2}!}\left[|\Psi_{1}|^{2}\left(\frac
{1+\eta}{2}\right)^{X_{1}}\left(\frac{1-\eta}{2}\right)^{X_{2}}+|\Psi_{2
}|^{2}\left(\frac{1-\eta}{2}\right)^{X_{1}}\left(\frac{1+\eta}{2}
\right)^{X_{2}}\right],
\end{equation}
with normalization
\begin{equation}
\sum_{X_{1}+X_{2}=X}P(X_{1},X_{2})=1.
\end{equation}
The diagonal elements of the density matrix are (see (24))
\begin{eqnarray}
p_{1}(X_{1},X_{2}) & = & \frac{|\Psi_{1}|^{2}(1+\eta)^{X_{1}}(1-\eta)^{X_{2}}}
{|\Psi_{1}|^{2}(1+\eta)^{X_{1}}(1-\eta)^{X_{2}}+|\Psi_{2}|^{2}(1-
\eta)^{X_{1}}(1+\eta)^{X_{2}}}\nonumber\\
p_{2}(X_{1},X_{2}) & = & \frac{|\Psi_{2}|^{2}(1-\eta)^{X_{1}}(1+\eta)^{X_{2}}}
{|\Psi_{1}|^{2}(1+\eta)^{X_{1}}(1-\eta)^{X_{2}}+|\Psi_{2}|^{2}(1-
\eta)^{X_{1}}(1+\eta)^{X_{2}}}\nonumber\\
p_{1}(X_{1},X_{2}) & + & p_{2}(X_{1},X_{2}) = 1.
\end{eqnarray}
The distribution (45) is a sum of two distributions with weights $|\Psi_1|^2$ and $|\Psi_2|^2$ and
with peaks at $X_1=X\frac{1+\eta}{2}$ and $X_1=X\frac{1-\eta}{2}$, respectively. The width of each
peak is $\frac{1}{2}\sqrt{X}$, and the distance between the peaks is $\eta X$. For given $\eta <<1$
and with $X\lesssim \frac{1}{4}\eta^{-2}$, the peaks are fused into one. With increasing $X$, they
begin to separate at $X\approx \frac{1}{4}\eta^{-2}$. Beyond this, the distance exceeds the width,
and for $X>>\eta^{-2}$, the two peaks become distinct.

We thus take the limit
\begin{equation}
X>>\eta^{-2}>>1,
\end{equation}
and introduce continuous variables,
$$
z=\frac{X_{1}-\frac{1}{2}X}{X\eta},~~~~~Z=2X\eta^{2};
$$
\begin{equation}
p_{j}(z)=p_{j}(X_{1},X_{2});~~~~q(z)~dz=P(X_{1},X_{2}),~~~~\int dz~q(z)=1.
\end{equation}
Here $z$ is a pointer variable, and $Z$ is a variable for the approach to
the thermodynamic limit of $M$. We then get
$$q(z)=q_{1}(z)+q_{2}(z),$$
$$q_{1}(z)=\sqrt{\frac{Z}{\pi}}~|\Psi_{1}|^{2}~e^{-Z(z-\frac{1}{2})^{2}},$$
\begin{equation}
q_{2}(z)=\sqrt{\frac{Z}{\pi}}~|\Psi_{2}|^{2}~e^{-Z(z+\frac{1}{2})^{2}},
\end{equation}
and
\begin{equation}
p_{1}(z)=\frac{q_{1}(z)}{q(z)},~~~~p_{2}(z)=\frac{q_{2}(z)}{q(z)},
\end{equation}
In the limit of $Z\longrightarrow\infty$,
$$
q_{1}(z)=|\Psi_{1}|^{2}~\delta\left(z-\frac{1}{2}\right),~~~~p_{1}\left(
+\frac{1}{2}\right)=1,~~~~p_{2}\left(+\frac{1}{2}\right)=0,
$$
\begin{equation}
q_{2}(z)=|\Psi_{2}|^{2}~\delta\left(z+\frac{1}{2}\right),~~~~p_{1}\left(-
\frac{1}{2}\right)=0,~~~~p_{2}\left(-\frac{1}{2}\right)=1,
\end{equation}
in agreement with the previous result (44).

To get an indication how fast the distribution $q(z)$ in (50) becomes bimodal with increasing $Z$,
we can follow the approach to zero of the
average of the geometric mean of the two probabilities $p_{1}$ and $p_{2}$,
$$
\langle\sqrt{p_{1}p_{2}}\rangle= \int_{-\infty}^{\infty}dz~q(z)
\sqrt{p_{1}(z)p_{2}(z)}=|\Psi_{1}||\Psi_{2}|\sqrt{\frac{Z}{\pi}}~
\int_{-\infty}^{\infty}dz~e^{-Z\left(z^{2}+\frac{1}{4}\right)}=
$$
\begin{equation}
=|\Psi_{1}||\Psi_{2}|~e^{-\frac{Z}{4}}=|\Psi_{1}||\Psi_{2}|~e^{-
\frac{1}{2}X\eta^{2}}.
\end{equation}
Now $X$ is the number of stochastic parameters and they can be chosen as
the number of components of the state of $M$, i.e., as the number of
dimensions of the Hilbert space of $M$. This grows exponentially with the
number of degrees of freedom, i.e., with the number of particles $N$
involved. Thus, the separation indicator (53) goes to zero with
growing $N$ like
\begin{equation}
\alpha~\mbox{exp}[-\beta \eta^{2}e^{\gamma N}],
\end{equation}
where $\alpha$,  $\beta$ and  $\gamma$ are positive parameters.

This means that the stochasticity necessary for the selection process
of a final state $-$ and a measurement result $-$ is provided by
relatively few particles in the beginning of the chain of interactions
within $M$. How correlations can develop when there are several such
chains will be briefly discussed in the next section.

\section{Correlations}

We go back to the same kind of model as in Section 6 but with only two
stochastic signs, $\varepsilon_{11}$ and $\varepsilon_{21}$ and with $\eta_{11}=\eta_{21}=\eta$,
i.e., only one step. Then the
distribution over ($\varepsilon_{11}, \varepsilon_{21}$)
corresponding to (45) is
\begin{equation}
P(\varepsilon_{11}, \varepsilon_{21})=\frac{1}{4}\left(1+\eta\left(|
\Psi_{1}|^{2}-|\Psi_{2}|^{2}\right)(\varepsilon_{11}+\varepsilon_{21})
+\eta^{2}\varepsilon_{11}\varepsilon_{21}\right).
\end{equation}
This clearly leads to a positive correlation between
$\varepsilon_{11}$ and $\varepsilon_{21}$,
\begin{equation}
\langle\varepsilon_{11}\varepsilon_{21}\rangle=\eta^{2}.
\end{equation}
Thus, entanglement of different, initially independent, parts of $M$,
with the quantum system $\mu$, leads to this kind of correlation through
their common influence on transition probability per unit time.

If we consider two sets of stochastic variables connected to
independent parts of the measurement apparatus $M$, such as different
detectors in an Einstein-Podolsky-Rosen experiment, and let them be
described (in the limit of many variables) by the continuous variables
$z$ and $u$ as in the previous section, then the equations corresponding to
(50) and (51) are
$$q(z,u)=q_{1}(z,u)+q_{2}(z,u),
$$
$$q_{1}(z,u)=\frac{\sqrt{ZU}}{\pi}~|\Psi_{1}|^{2}~e^{-Z(z-\frac{1}{2})^{2}}
~e^{-U(u-\frac{1}{2})^{2}},
$$
\begin{equation}
q_{2}(z,u)=\frac{\sqrt{ZU}}{\pi}~|\Psi_{2}|^{2}~e^{-Z(z+\frac{1}{2})^{2}}
~e^{-U(u+\frac{1}{2})^{2}}
\end{equation}
and
\begin{equation}
p_{1}(z,u)=\frac{q_{1}(z,u)}{q(z,u)};~~~~p_{2}(z,u)=\frac{q_{2}(z,u)}{q(z,u)}.
\end{equation}
In the limit $Z,U\longrightarrow\infty$, this implies total
correlation between the pointer
variables $z$ and $u$.

\section{A Darwinian perspective}

One can look upon the measurement process, i.e., the mapping of the out state of the scattering
process within  $\mu$, on the final state, including registration of the measurement result, as an
evolution process, taking place with the density matrix for final states as the relevant phenotype.
There is no external influence (from outside $\mu$ and $M$) needed to describe this process:
scattering + measurement is to be viewed as an integrated process.

It should be emphasized that the out state of the scattering process itself is \textit{not} realized
in the actual case due to the interaction between the microsystem $\mu$ and the measurement
apparatus $M$. The mapping is thus a mapping from a situation \textit{without} $M$ to a situation
\textit{with} $M$ present. The whole evolution process represents one single measurement.

The replicators in the evolution process are the individuals (the final states), each represented by
its genetic code, for the $X$th generation,
$(\underline{\varepsilon}_{1},\underline{\varepsilon}_{2},...,\underline{\varepsilon}_{X})$.

The genetic information is inherited from one generation to the next without any mutation. In the
replication from the $(X-1)$th generation to the $X$th generation, new genetic material
$\underline{\varepsilon}_{X}$ is added, bringing in new variability. The phenotype of interest is
described by the corresponding generation of the density matrix, reduced to $\mu$,
$\rho_{\mu fX}=\mbox{Tr}_{M}[\rho_{fX}]$. The fitness is described by
$\sum_{l=1}^{n}|\Psi_{l}|^{2}B_{l}(\underline{\varepsilon}_{X})$.

In the evolution, the change in phenotype in the $X$th generation is described by (30). The fitness,
normalized to a probability distribution over the additional genetic information (31), is given by (
32). This leads to a drift and a random walk as described by (33). The non-diagonal elements drift
towards zero, step by step, and the diagonal elements take part in a random walk without drift as
described by (35),with the corners (the eigenstates of the observable) as attractors.

\section{Conclusions}

The result just obtained means that the interaction between $\mu$ and $M$ has
the characteristics of a measurement process with $\mu$ being measured upon
by means of $M$.

In the final transition matrix, there is a non-linear dependence on
the partial transition amplitude for $\mu$ in isolation as described in
Section 2 (the modulus
squared of the "wave function" $\left.|\Psi_{j}|^{2}\right)$.
The non-linearity
in $|\Psi_{j}|^{2}$ (in (24), (29)
or (35)) comes from the mixing of the two kinds of interaction: the
internal interaction within $\mu$, and the final-state interaction
between $\mu$
and $M$. The consequence of abandoning the assumption X,
referred to in the introduction, is to treat these
two interactions as one whole within scattering theory.
The largely
unknown state of $M$ was brought in through the stochastic variables
$\underline{\varepsilon}.$ This was done stepwise through the mapping
procedure of Section
5. \textit{At no point was
there a
break away from quantum mechanics}, here applied in the form of quantum
scattering theory. The property of $M$ to be unbiased was brought in
through the
factor (21) together with the assumption of an even à priori initial-state
probability distribution over $\underline{\varepsilon}.$

Correlations between the stochastic variables of $M$, related through
entanglement with $\mu$, build up, as described in Section 7. This is
clearly essential for an analysis of causality concepts.

The mechanism presented here for the selection process in the $\mu$-$M$
interaction also opens new questions. One is what the process looks
like in space and time. Here this question
is hidden behind scattering theory involving infinite time.
Another is the question indicated above, how a useful concept of
causality could
be defined and applied. A third question is how to understand in more
detail the
unbiased metastable state of $M$ and the origin of the factor (21).

The present work is in the tradition of references \cite{2} to \cite{5} and \cite{8}.
The ambition is to let the density matrix for the final state
describe the result of a single measurement and not only the ensemble of measurements. The
random walk connected to this is characterized by the changes of the density
matrix for $\mu$, given by (33) and, for the diagonal elements, by (35). The
first equation (33) leads to a decoherence for the ensemble of density
matrices
through a drift of the non-diagonal elements to zero; the second leads to a
random walk towards the corners of the probability simplex, i.e., to a reduction
of the wave function.

Mathematically, the mechanism descibed here is very close to the diffusion
process of Gisin and Percival, but instead of time, we use system
extension, and
we stay within linear quantum mechanics. However a space-time analysis of the process
suggested here would probably lead to a diffusion process of the kind
that
Gisin
and Percival \cite{5} have suggested.

Finally, a small comment on the ambitions of Einstein and Bohr
revealed in their
discussion on measurement in quantum mechanics.

Einstein did not like the idea that God plays dice behind the scene. He
might have accepted a dice-throwing that can be explained in a
statistical way.
Here the source of randomness is not lack of microcausality but
stochasticity of
an object with many unknown degrees of freedom. Through entanglement and
enhancement/inhibition, this stochasticity leads to correlation build-up and
finally to a total bifurcation into different final states of $\mu$, the
eigenstates
of the measured observable $A$.

Niels Bohr emphasized the necessity of having a measuring instrument of a
classical nature. Here the selection process (reduction of the wave packet)
takes place only in the thermodynamical (classical) limit of the system $M$.

In the language of Dennett \cite{9}, the \textit{cranes} of quantum
mechanics seem adequate
for analysing the measurement process, and no \textit{skyhook}, such
as a randomness
without an identifiable source or a many-world interpretation or a non-linear extension of
quantum mechanics, seems to be needed.

\vspace{2 cm}


\noindent
\large{\textbf{Acknowledgements}}

\noindent
\normalsize{
Most of this work was done during my time at the Department of Physical Resource Theory at Chalmers
University of Technology and G\"{o}teborg University, and I thank my colleagues there for a creative
research environment. I thank Professors Bengt Månsson, Steven Hwang and Claes Uggla for hospitality
at my new work place. Part of the present work was also done during stays at the National Centre for
Mathematical Sciences in Accra and the Department of Physics of the University of Cape Coast, and I
thank Professors F.K.A. Allotey, S.K. Adjepong and S. Yeboah-Mensah for their kind hospitality.

I thank Maria Grahn and Jens Fjelstad for their kind assistance in making the paper ready for this
kind of publication.

Over the years, I have troubled many colleagues with discussions on ideas at various stages of
development. Let me mention here Francis Allotey, Predrag Cvitanovic, GianCarlo Ghirardi, Nicolas
Gisin, Hermann Haken, Hans Hansson, Kristian Lindgren, Martin Nilsson, Kazimierz Rzazewski and
Franco Selleri. I thank them all for their interest and patience.

The work of Gisin and that of Gisin and Percival together has been a challenging source of
inspiration for me. The constant encouragement given to me by Kristian Lindgren and Magdalena
Eriksson has also been very important for me.
}

\pagebreak

\vspace{6 cm}
\pagebreak
\thispagestyle{empty}

\noindent
\large{\textbf{Appendix 1: Soft photon exchange}}\newline

\normalsize{}
\noindent
Soft-photon exchange and emission is an old and well-known example of
factorizable processes in quantum electrodynamics. When this was
properly understood, the picture of scattering was drastically changed
in the sense that no non-forward scattering of charged particles takes
place without soft-photon emission.
Later, this was identified as coherent radiation from a classical
current describing the charged particles.

Here we choose a simple case to show the factorization of soft photon emission
or exchange. We consider an outgoing electron (charge $-e$, mass $m$) with
final momentum $p$, described by a spinor $\overline{u}(p)$,
$$
p^{2}=m^{2};~~~ \overline{u}(p)(ip\cdot\gamma+m)=0~~~~~~~~~~~~~~~~~~~~~~~~~~~~~~~~~~~~~~~~~~~~~~~~~~
~~~~~~~~~~~~~~~~~~~~~~~~
  (A1.1)
$$
after emitting two soft photons with momenta $k_{1},k_{2}$  and
polarizations $e_{1},e_{2}$,
$$
\begin{array}{lr}
k_{1}^{2}=k_{2}^{2}=0;~~~k_{1}\cdot e_{1}=k_{2}\cdot e_{2}=0;&\\
|\mathbf{k}_{1}|,|\mathbf{k}_{2}|<<m ~~~~~~~~~~~~~~~~~~~~~~~~~~~~~~~~~~~~~~~~~~~~~~~~~~~~~~~~~~~~~~~
~~~~~~~~~~~~~~~~~~~~~~~~~~~~~~~~~~~~
& (A1.2)
\end{array}
$$
In the evaluation of the Feynman diagram (Fig. 1), the spinor
$\overline{u}(p)$ is then
changed into a term proportional to
$$e^{2}\overline{u}(p)\left[e_{1}\cdot \gamma\frac{i(p+k_{1})\cdot \gamma+m}
{(p+k_{1})^{2}+m^{2}}e_{2}\cdot \gamma+(1\leftrightarrow
2)\right]~\frac{i(p+k_{1}+k_{2})\cdot \gamma+m}{(p+k_{1}+k_{2})^{2}+
m^{2}}=~~~~~~~~~~~~~
$$
$$
=e^{2}\frac{1}{2(p\cdot k_{1}+p\cdot
k_{2})}~\overline{u}(p)\left[\frac{e_{1}\cdot \gamma(ip\cdot \gamma+m)e_{2}\cdot
\gamma(ip\cdot \gamma+m)}{2p\cdot k_{1}}+(1\leftrightarrow
2)\right]~~~~
  (A1.3)
$$
$$=e^{2}\frac{-p\cdot e_{1}p\cdot e_{2}}{p\cdot k_{1}+p\cdot k_{2}}\left(\frac{1}{p\cdot k_{1}}+
\frac{1}{p\cdot k_{2}}\right)\overline{u}(p)=(s(k_{1})\cdot e_{1})(s(k_{2})\cdot e_{2})
\overline{u}(p),~~~~~~~~~~~~~~~~~~~~
$$
where
$$
s_{\mu}(k)=-e \frac{ip_{\mu}}{p\cdot k}=-e\int_{0}^{\infty}dt\int d^{3}xe^{i
\left(\mathbf{k}\cdot \mathbf{x}-|\mathbf{k}|t\right)}~\delta^{3}
\left(\mathbf{x}-\frac{\mathbf{p}}{p_{0}}t\right)
\frac{p_{\mu}}{p_{0}}.~~~~~~~~~~~~~~~~~~~(A1.4)
$$
is the Fourier transform of the current of a classical point
charge $-e$ moving
from $\mathbf{x}$=0 at time zero with velocity $\mathbf{p}/p_{0}$. The rest of the diagram is
unchanged in the limit of small $k_{1},k_{2}$.

Equation (A1.3) states that the emission of the two photons is
described by one
independent scalar emission factor for each photon. The corresponding
holds for
two photons being absorbed by the electron.

For $m$ photons, use is made of the identity,
$$
\sum_{(i_{1}i_{2}...i_{m})}\frac{1}{a_{i_{1}}\left(a_{i_{1}}+
a_{i_{2}}\right)...\left(a_{i_{1}}+a_{i_{2}}+...+a_{i_{m}}\right)}=
\frac{1}{a_{1}a_{2}...a_{m}}.
~~~~~~~~~~~~~~~~~~~~~~~~~~~~~(A1.5)
$$
There is also a factor $(m!)^{-1}$. Summation over photon states and
over $m$, then
gives rise to a coherent state generated by the classical current
(A1.4). Different external particles contribute independent factors.

This is our basis for changing (8) into (15).\newline

\noindent
\textbf{Remark A1.1}: The non-linear terms of the kind deriving from the
denominator in (24) were not treated in the old literature of soft
radiation. The reason was that these terms did not contribute to the
infrared divergences, i.e., the divergences connected to the limit of
zero photon mass, and were therefore not needed for regularizing the
theory.

The dependence of the transition probabilities on the detailed
geometry of soft photon detection is not very strong.
Therefore soft photon radiation did not
immediately suggest non-linearities that could otherwise have shown a way to
handle the measurement process.\newline

\noindent
\textbf{Remark A1.2}: It is possible to show that the inclusion of soft photon
emission leads to decoherence in the sense that the non-diagonal
elements of the density matrix $\rho_{f}$ disappear. If the time
development is
analysed, one can see that this is a very slow process due to the
relative weakness of the electromagnetic interaction. In the case of a
soft electromagnetic field coupled to measurement device $M$, the large
number of degrees of freedom available in $M$ compensates for this
restriction. (Moreover, the stochastic interaction within $M$ leads to
the selection process, described by a bifurcating random walk,
resulting in a definite measuring result.)

\begin{center}
\begin{figure}[H]
\scalebox{0.9}{\includegraphics{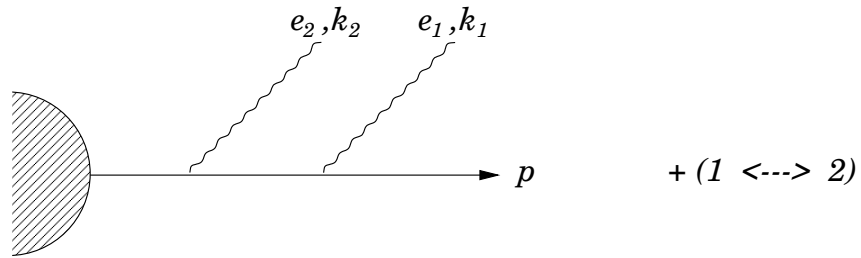}}
\caption{Feynman diagrams corresponding to $(A1.3)$ for emission of
two photons
  characterized by polarizations and momenta, $e_{1},k_{1}$ and
$e_{2},k_{2}$.}
\end{figure}
\end{center}



\vspace{13 cm}
\thispagestyle{empty}
\pagebreak
\thispagestyle{empty}

\noindent
\large{\textbf{Appendix 2: Non-linearity in "wave-function" $\Psi$
of overall S-matrix}}\newline

\normalsize{}
\noindent
The amplitude for a transition of $\mu$ from the initial state
$|0\rangle_{\mu}$ to an outgoing state $|j\rangle_{\mu}$
can be illustrated by the diagram $a$ of
Fig. 2. Here $\mu$ is represented by two particles in the ingoing and
the outgoing
states, and the scattering amplitude for the transition within $\mu$ is
represented
by a shaded circle. The interaction between $\mu$ and $M$ taking place
in the outgoing
state is included in diagram $b$ of Fig. 2 for the modified transition
amplitude
and represented by double lines.

The normalized total S-matrix element is obtained through diagrams to
all orders
of mixed interactions ($S$ and $S^{\dagger}$ within $\mu$ together with
$\mu$-$M$  interaction) as in diagram
$c$. This gives rise to a non-linear dependence on the internal transitions
amplitudes $\Psi_{j}$ for $\mu$, here playing the role of wave function
components.

\begin{center}
\begin{figure}[H]
\scalebox{0.7}{\includegraphics{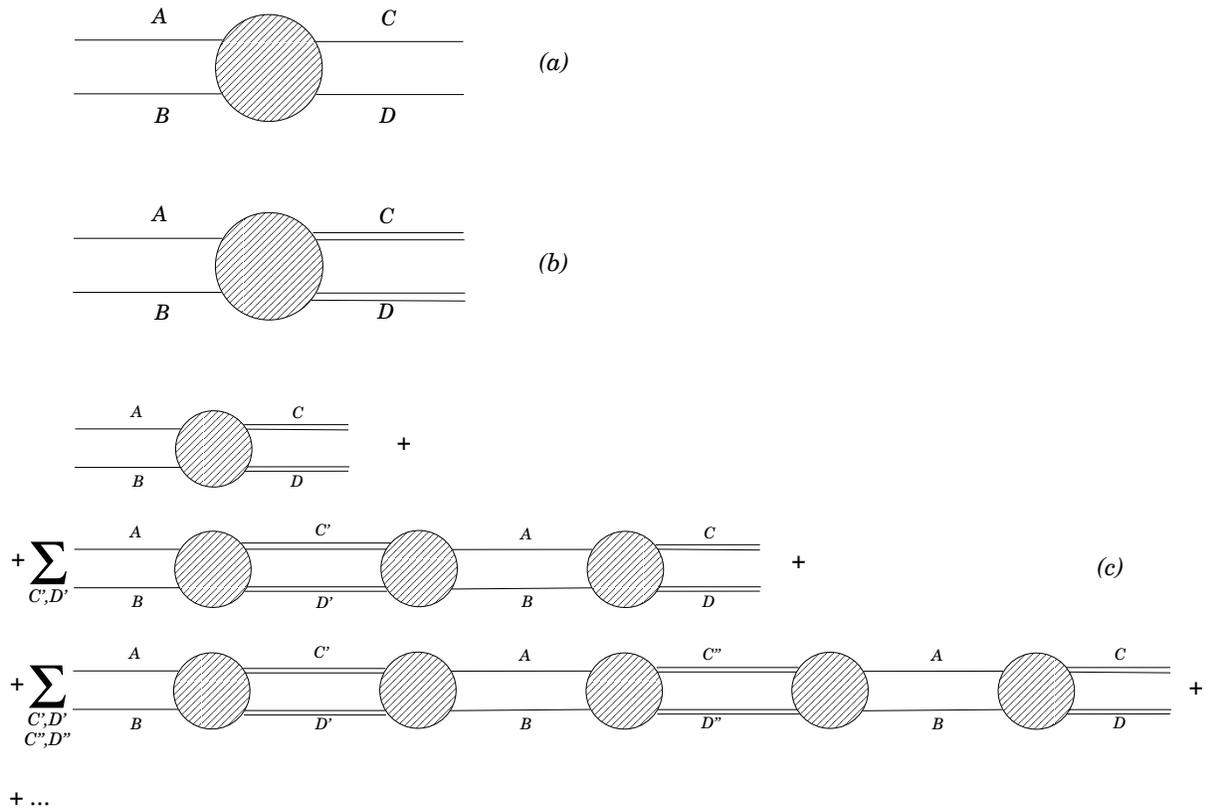}}
\caption{S-matrix diagrams for scattering in $\mu$,
$A+B\longrightarrow C+D$, without $M$ (diagram $a$) and with $C$ and
$D$ interacting with $M$ (diagram $b$). The sum over diagrams to all
orders (as indicated in $c$) should be taken into account.}

\end{figure}
\end{center}



\end{document}